\documentclass{aa}
\usepackage{epsfig} 
\usepackage{amsmath}
\usepackage{amssymb}
\usepackage{natbib}
\usepackage{multirow}
\usepackage{rotating}
\usepackage{epic,ecltree,shadow,fancybox}

\newcommand{\curl}{\operatorname{curl}}
\newcommand{\dv}{\operatorname{div}}

\newcommand{\lapl}{\operatorname{\Delta}}

\newcommand{\ovec}[1]{{\mbox{\boldmath $#1$}}}
\newcommand{\tmax}{{\text{max}}}
\newcommand{\Bvec}{\ovec{B}}

\newcommand{\vvec}{\ovec{v}}

\newcommand{\Rm}{\operatorname{\mathit{Rm}}}

\newcommand{\tcrit}{{\text{crit}}}

\newcommand{\tpol}{{\text{pol}}}
\newcommand{\ttor}{{\text{tor}}}

\newcommand{\myref}[1]{~\hspace{0pt plus 1pt minus 1pt}\ref{#1}}

\newcommand{\figsandref}[2]{Figs.\myref{#1} and\myref{#2}}
\newcommand{\figsref}[2]{\figsandref{#1}{#2}}

\newcommand{\figref}[1]{Fig.\myref{#1}}
\newcommand{\tablref}[1]{Table\myref{#1}}
\newcounter{saveqn}
\newcommand{\alpheqn}{\refstepcounter{equation}\setcounter{saveqn}{\value{equati
on}}%
\setcounter{equation}{0}%
\renewcommand{\theequation}{\mbox{\arabic{chapter}.\arabic{saveqn}\alph{equation
}}}}
 
\newcommand{\reseteqn}{\setcounter{equation}{\value{saveqn}}%
\renewcommand{\theequation}{\arabic{chapter}.\arabic{equation}}}

\newcommand{\tmerid}{{\text{merid}}}
\newcommand{\tdom}{{\text{dom}}}

\newcommand{\Emag}{E_{\text{mag}}}

\begin{document} 
 
\titlerunning{Proto--Neutron--Star Dynamos}
\date{Received \today / Accepted \today}
% \thesaurus{08.14.1;   % Stars: neutron 
%              08.16.6;   % (Stars:) pulsars: general 
%              02.13.1    % Magnetic fields 
%             } 
 
\title{The proto--neutron--star dynamo --- viability and impediments} 

\author{M. Rheinhardt\inst{1} \and U. Geppert\inst{2}}  
\institute{Astrophysikalisches Institut Potsdam,
 An der Sternwarte  16, 14482 Potsdam, Germany, \email{MReinhardt@aip.de}
 \and
 Max-Planck-Institut f\"ur Extraterrestrische Physik, Gie{\ss}enbachstra{\ss}e,
 85748 Garching, Germany, \email{urme@xray.mpe.mpg.de}}
\offprints{M. Rheinhardt}

\date{Received date ; accepted date} 
 
\abstract{We study convective motions
taken from hydrodynamic simulations of rotating proto--neutron stars (PNSs)
with respect to their
ability to excite a dynamo instability which may be responsible for the
giant neutron star magnetic fields. Since it is impossible to simulate the
magnetic field
evolution employing the actual magnetic Reynolds numbers ($\Rm$) resulting from
the hydrodynamic simulations, (smallest) critical $\Rm$s  
% Both different regions and
%evolutionary phases of the proto- neutron star are investigated. By use of a
and the
corresponding field geometries are derived on the kinematic level by rescaling the
velocity amplitudes.
It turns out that the actual values of $\Rm$
are by many orders of magnitude larger than the critical values found.
A dynamo might therefore start to act vigorously
very soon after the onset of convection. But as in general dynamo growth
rates are non--monotonous functions of $\Rm$ the later fate of the magnetic
field is uncertain.
Hence, no reliable statements on the
existence and efficiency of PNS dynamos can be drawn without considering the interplay
of magnetic field and convection from the beginning. 
Likewise, in so far as convection inside the PNS is regarded to be essential
in re--launching the supernova 
explosion, a revision of its role in this respect could turn out to be necessary.

}
 
\maketitle 

\section{Introduction}
The origin of the neutron star (NS) magnetic fields is subject of scientific 
debate since the discovery that the pulsar mechanism is based on the existence
of the strongest magnetic fields seen in the universe.
Though the idea of magnetic flux conservation during the
collapse of the NS progenitor came up quite naturally, doubts as to whether that mechanism explains the
observed field strengths were raised almost simultaneously \citep[see, e.g.,][]{TD93}.
That macroscopic currents -- as almost everywhere else in the universe -- are the
cause of the NSs' magnetic fields seemed obvious; thus attention has been
drawn to the idea of a dynamo.
\cite{FR77} first pointed out that such an instability can start to act
immediately after the neutron star's birth, and last until its internal
motions disappear. 

The increasing understanding of the supernova mechanism  
\citep[e.g.,][and references therein]{BF92} and the accompanying
process of NS creation strengthened the idea of a convective proto--neutron star
(PNS) phase. This very first epoch in a NS's life starts immediately
after the collapse when negative radial
gradients of both entropy and lepton number are created, thus in general
enabling Ledoux convection (see, e.g.,
\cite{E79} and \cite{BL86}) and lasts about half a minute.
\cite{TD93} sketched 
the possibility of dynamo action caused by turbulent convective motions
in the PNS. They claimed that as these
vigorous motions combine with differential rotation, a mean--field dynamo of $\alpha$--$\Omega$ type
may act in regions where the rotation is capable of influencing the convection
significantly. By use of a mixing--length approach
they estimated
the convective velocity to be about $10^8$ cm s$^{-1}$ and concluded from equipartition
that such a dynamo may
amplify a seed field up to $10^{15}$G during the first $30$ seconds of a NS's life. In contrast,
small--scale dynamo action without generation of a coherent
large--scale field should be expected where the influence of the rotation on the convection is
negligible.

In two complementary papers, \cite{BRU03} and \cite{UG04} recently
presented studies on field generation in
PNSs where the former deals with mean--field dynamos and the latter
with small--scale ones. 
%both possibilities already figured out by \cite{TD93} are more thoroughly studied

In their strongly simplified model, \cite{BRU03} assume the mean e.m.f. to be 
constituted by isotropic $\alpha$ and $\beta$ effects only
and adopt purely radial dependences for 
differential rotation,
the $\alpha$--parameter and the turbulent magnetic diffusivity. 
Like \cite{TD93} and based on the results of \cite{MPU02},
they argue that $\alpha$ is negligible in an inner, entropy--gradient driven
convective region, but proportional to the rotation rate $\Omega$ 
in an overlaying, lepton--gradient driven convective shell.
They derive critical values of $\Omega$ above which dynamo action in 
either the $\alpha^2$ or the $\alpha$--$\Omega$ regime is possible
and conclude that the vast majority of PNSs will show mean--field dynamo
action.  
%for rotational periods very likely for PNSs a dynamo will act.
%While in the inner core the convective velocity is assumed to be in the order of
%$10^8$ cm s$^{-1}$, in the outer shell it has the same order of magnitude as
%found by \cite{FH00}. 
%not taking into
%account already available velocity profiles from realistic convection
%simulations.

The work of \cite{UG04} is focused on small--scale dynamos which are supposed to emerge from both
convective regions. As the magnetic Reynolds numbers are huge even when calculated
with the scales of the convective eddies, at least a necessary condition for this supposition
is surely satisfied. 
%This seems reasonable in view of the huge magnetic Reynolds numbers which
%result even when referring to the scales of the convective eddies.
To get estimates for the magnetic field strength at the end of the convective
phase, equipartition is employed thus arriving at
magnetic field strengths of $\approx 10^{13}$G in both convective regions for the largest turbulent
scales $\approx 1\,\ldots\,3$ km.

Both studies state that given the estimated convection amplitudes, dynamos can act because
the velocities 
are overcritical with respect to these dynamos.
\cite{UG04} do not estimate the real level of `overcriticality', i.e. the `distance' between the real 
and the critical velocity (e.g. as a ratio of amplitudes) at all.
\cite{BRU03} provide a hint that in terms of the mean--field coefficient $\alpha$ the real flow might be
overcritical by a factor up to 1000 but refrain from discussing the consequences on the basis of the behavior of growth rates.

But, as in general the growth rates do not simply grow with growing
overcriticality (in the above sense) no reliable conclusion on dynamo action
can be drawn without calculating them based on realistic velocity
amplitudes. 
Moreover, it appears desirable to overcome the drawback of using
ad--hoc assumptions on the convection and instead to  make use of `realistic' convective velocity patterns obtained from hydrodynamic
simulations of the PNS stage. By feeding them into dynamo calculations `as they
are', a better
understanding of the different PNS layers' contributions to the field generation and hence the relative importance of 
small and large--scale dynamos could be gained.        
  
Along with the problem of magnetic field generation, the comparatively short period of
convection in PNSs has been taken to play an important role in
the puzzle of another major question: How can a sufficiently large amount of heat and/or
neutrinos be transported into the region behind the shock front to enable the
re--energized supernova explosion?

Recently, doubts have been reinforced about whether the PNS convection alone is sufficient
to accomplish this and new hints have been provided that convection in a second outer spherical shell (50 \!\ldots 100 km)
could be crucial instead \citep[see][and references therein]{J04a}.
\footnote{It seems conceivable that a magnetic field is being generated in this outer zone, too, parts of which could later constitute
the crustal magnetic field by virtue of fallback accretion.}
But at least in the case when the thermal energy stored
outside the neutrinosphere and transported by this convection towards the shock is not sufficient
to re--launch it, the transport of neutrinos trapped inside the neutrinosphere by virtue of the coupled
action of the inner (PNS) and outer convective zones could help: If the overshooting zone of the
former and the 
undershooting zone of the latter overlap, a sufficient number of neutrinos would be enabled to escape 
from the PNS and to reach the post-shock heating zone.

Up to now, all attempts to cope with the problem of the convection--driven explosion
were made with the influence of the magnetic field neglected.
But, if success or failure 
depends sensitively on details of the convective phase, a clear understanding of
how early and how fast the magnetic field starts to act back on the motions 
which generated
it (perhaps in both convection zones) seems to be essential.
\cite{MPU02} were the first to include the influence of an imposed magnetic field
on the conditions for the onset 
of the relevant convective instabilities and found that only extremely strong
($10^{15}$G) magnetic fields suppress the occurrence of convection completely.

Since the mid--nineties 
fully nonlinear hydrodynamic supernova simulations also resolving the PNS started to provide
more detailed insight into the convection there \citep[see, e.g.,][]{K97,JK98,FH00,J04,J04a}.

%Another series of papers \citep{MPU00,MPU02,MPU04} deals with the conditions for the onset 
%of the relevant convective instabilities by means of a linear analysis.

\section{Kinematic dynamo action of PNS convection}
The purpose of this paper is to study the dynamo--related properties of velocity profiles found in hydrodynamic PNS simulations.
To start with, we restrict ourselves to the kinematic approach, that is, we solve the induction equation with
prescribed velocities $\vvec$. It reads in
normalized form for a homogeneous medium
\begin{equation}
\dot{\Bvec} = \lapl\Bvec + \Rm\curl(\vvec\times\Bvec)\,,\quad
\dv \Bvec = 0
\label{indeqlin}
\end{equation}
where $\Rm=UR/\eta$ is the magnetic Reynolds number with a characteristic
velocity $U$, the magnetic diffusivity $\eta$ and the radius $R$ of the PNS.
We define $\vvec$ by snapshots
\footnote{Using the originally time--changing convection instead of snapshots   
would not influence our conclusions qualitatively.}
taken from the
simulations of \cite{K97} \citep[see also][]{JK98}
who performed 2D (axisymmetric) simulations of rotating PNS convection
with a model including a completely radial neutrino transport scheme and employing the EOS of Lattimer and Swesty
without accretion. To check both an early stage possibly relevant for the explosion  and a stage with fully developed
convection perhaps determining the later NS field, we took the velocity profiles at $30$ ms and $0.9$ s after bounce.

\begin{figure*}
\hspace*{-0mm}\begin{tabular}{c@{\hspace*{1cm}}c}
\epsfig{file=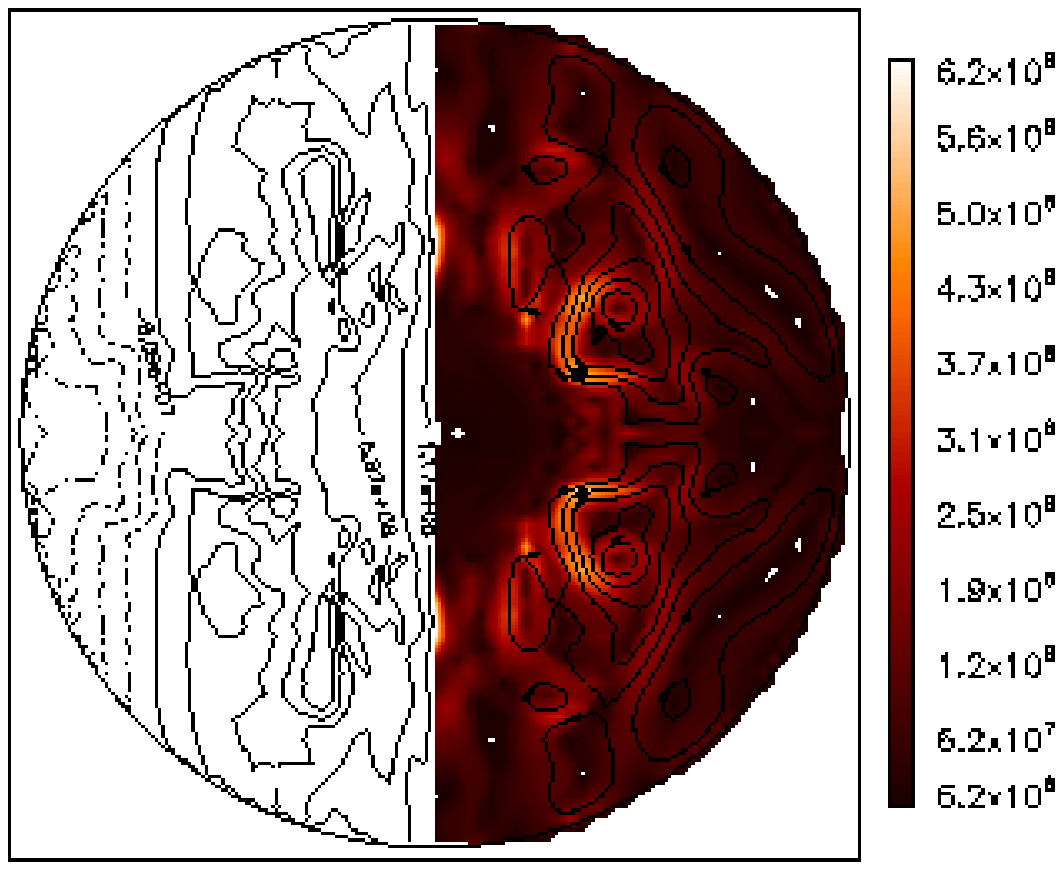, width = 0.45\linewidth}%
%
%\epsfig{file=/home/rei/mhdu/nsuk9_um_z_0.0000.ps, width=.5\linewidth}%
%\epsfig{file=/home/rei/mhdu/nsuk9_ub_z_0.0000.ps, width=.5\linewidth}%
%\epsfig{file=/home/rei/mhdu/nsuk9_uu_z_0.0000.ps, width=.5\linewidth} &
%\hspace*{-.5\linewidth}%
 &
\epsfig{file=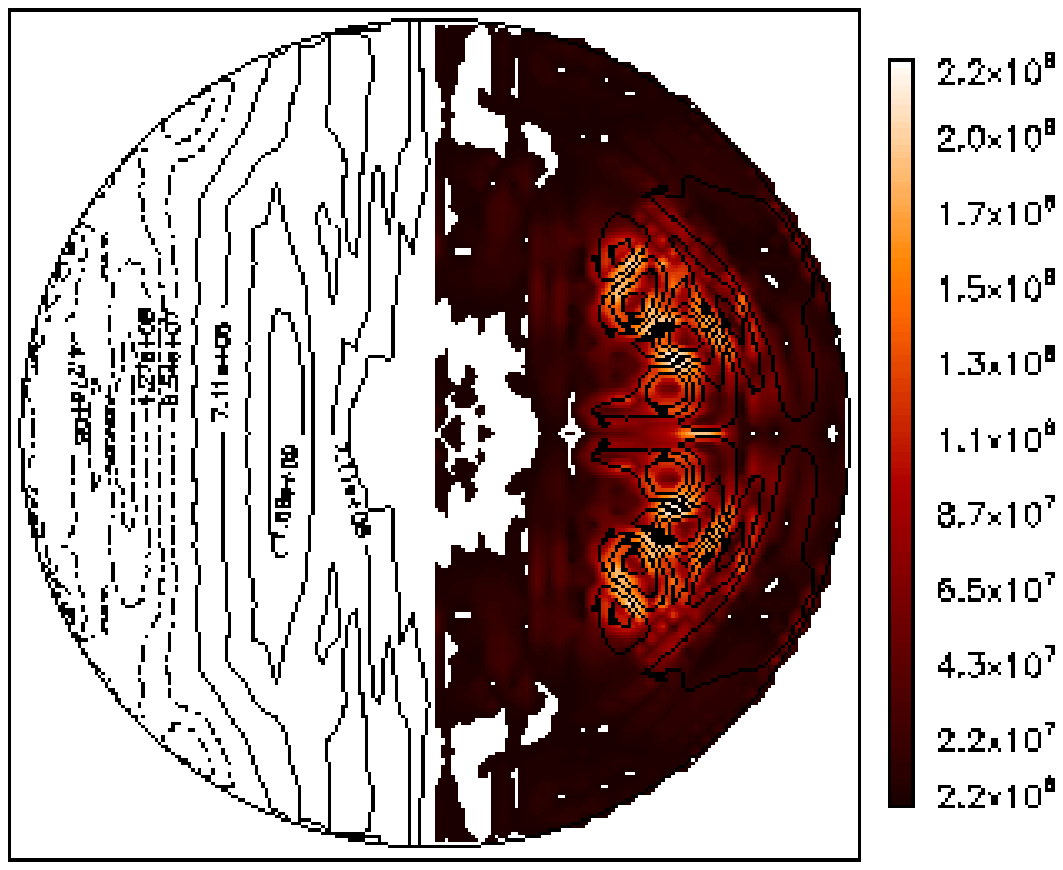, width = 0.45\linewidth}%
%\hspace*{-.5\linewidth}%
%\epsfig{file=/home/rei/mhdu/nsuk2_uh_z_0.0000.ps, width=.5\linewidth}%
%
%
\end{tabular}
\caption{\label{janv} Convective velocities simulated by \cite{K97} for
$30$ ms
(left) and $0.9$ s (right) after bounce in cm s$^{-1}$. In the left semicircles the
isolines of the azimuthal velocity and in the right ones the streamlines of the 
meridional velocity are shown. Note that the former is measured
with respect to a frame in which the flow has zero angular momentum.
%(see Table \ref{vprop}).
In the right semicircles background colors provide 
the modulus of the meridional velocity.
} 
\end{figure*}
The geometrical structures of the convective velocities $\vec{v}$ are shown in 
\figref{janv};
Table \ref{vprop} summarizes their main qualitative features.
At the late stage the azimuthal velocity component (i.e., the differential
rotation) is in a
large portion of the PNS's volume nearly independent
of $z$, as a consequence of the Taylor--Proudman theorem.
The early velocity pattern is
dominated by only a few convective cells per hemisphere whereas the late one
shows more and, hence, smaller cells.    

\begin{table}
\caption{\label{vprop} Qualitative features of the convective
velocities: $R$ - PNS radius, $R_{\text{i}}, R_{\text{o}}$ - estimated inner and outer radii of the convective shell;
$N_\theta$, $N_r$ - number of prominent convective cells in meridional and
radial directions; $v_\tmax$ - maximum convective, $v^\tmerid_\tmax$ - maximum
meridional velocity.
%$\Omega$ - angular velocity
%of a rigidly rotating sphere with the same angular momentum (constant density assumed).
$\Rm$ - magnetic Reynolds
number based on the r.m.s. value of the velocity with respect to the full PNS volume and an electric conductivity 
of $10^{24}\mbox{s}^{-1}$.
}
\begin{center}
\begin{tabular}{l@{\hspace{5mm}}cc}
		      \hline					    
		      \hline\\[-.6cm]			     \\
                     	&   30 ms 	&   0.9 s    \\
		      \hline			     \\
$R$ (km)		&  59 		&   22       \\
$R_{\text{i}}/R$      	&  0.3   	&   0.5	     \\
$R_{\text{o}}/R$      	&  0.75  	&   0.78     \\
$N_\theta$ 		&   2		&    8       \\
$N_r$      		&   1     	&    2       \\
$v_\tmax$ (cm/s)       	& $1.3\cdot 10^9$ & $2.2\cdot 10^9$          \\
$v^\tmerid_\tmax$(cm/s)& $6.3\cdot 10^8$ & $2.2\cdot 10^8$         \\
%$\Omega$ (1/s)		& 240		&  1360       %\\ 
$\Rm$           	& $4.2\cdot 10^{19}$ &  $2.2\cdot 10^{19}$       
\end{tabular}
\end{center}

\vspace{-.4cm}
\end{table}
As the velocity patterns are reflectionally
symmetric with respect to the equatorial plane and symmetric about the PNS's axis of rotation, any eigensolution of \eqref{indeqlin}
is either equatorially symmetric (S) or antisymmetric (A) and its spherical components vary like
$\cos m\varphi$ or $\sin m\varphi$, $m=0,1,2,\ldots$ where $\varphi$ is the azimuth with
respect to the rotation axis. For non--decaying solutions the case $m=0$ (axisymmetry) is excluded by Cowling's theorem.
As a consequence, if dipolar dynamo solutions, i.e. in our case, S1--solutions, occur their axes are bound to lie in the equatorial plane.
Note that this restriction is only
due to the axisymmetry of $\vvec$ and does not apply to a realistic situation with 3D
convection. Then, it is well possible that a dominating axisymmetric (aligned) dipolar field, accompanied
by non--axisymmetric constituents is generated.   
We restrict ourselves here to the case $m=1$ only as from experience we expect those modes
to be the most easily excitable ones. Both equatorial symmetries are included, that is,
for either velocity the A1 and S1 solutions are considered.

For the boundary conditions at the surface we assumed the surroundings of the PNS to exhibit a significantly
lower electric conductivity than its interior. We modeled this condition by requiring that the magnetic field
is at the surface continued into a vacuum field.

As we cannot hope to solve Eq. \eqref{indeqlin} with the velocity snapshots' 
real values of $\Rm$ (see \tablref{vprop}) we restrict ourselves instead to determining
the so--called {\em critical values} of $\Rm$, $\Rm_\tcrit$, for which the total magnetic field energy
(or a suitable temporal average of it) neither
grows nor decays. The corresponding magnetic field solution is then called {\em marginal}.
To accomplish this, the magnitude of the velocity was considered freely
rescalable while its geometry remained fixed. 
That is, we simply multiplied the complete velocity field with a factor which was adjusted such that a marginal
solution emerged.   
In mathematical terms, we sought solutions of Eq. \eqref{indeqlin} with an exponential ansatz for the time
dependence of $\Bvec$, i.e., $\partial/\partial
t=\lambda$, where the real part of the time increment $\lambda$ has to vanish. %$,\,\Re\{\lambda\}=0,$ 
In general, this is not equivalent with stationarity of the marginal solutions as a non-vanishing imaginary part
of $\lambda$ may occur causing oscillating marginal solutions. In our situation, however, 
in which we excluded axisymmetric ($m=0$) solutions, the oscillations introduced in this way are restricted to
uniform rotations of the field pattern about
the axis of rotation, where the imaginary part $\Im\{\lambda\}$ defines the rotation rate. Hence, the magnetic field energy is 
for any marginal solution a constant quantity.

Our actual technique to determine $\Rm_\tcrit$ together with the corresponding marginal solution
was to integrate \eqref{indeqlin} as an initial value problem. We started with values of $\Bvec$ specified according to
the desired symmetry type (A1 or S1) of the solution, but arbitrary otherwise.
In every time step we adjusted $\Rm$ (automatically) such that the total magnetic energy approached a
stationary value. As no marginal solutions exist for which the total magnetic
energy is oscillating no modes of a given symmetry type can be overlooked 
the $\Rm_\tcrit$ of which could possibly be smaller  than the
detected one.
%The modification  of \eqref{indeqlin} consists in
%adding an artificial `quenching' factor $1/(1+\epsilon E_B)$ in front of the induction term $\uvec\times\Bvec$
%or, what is the same, in incorporating this factor in $\Rm$
%where $E_B$ is the total magnetic field energy and $\epsilon$ a positive number determined by numerical needs
%only. This measure ensures, that for overcritical values of $\Rm$, i.e., $\Rm>\Rm_\tcrit$, when the magnetic energy 
%grows, $\Rm$ is reduced while it is enlarged in the opposite case. With a
%suitably chosen $\epsilon$
%this procedure quickly converges and yields a stationary solution which is just the marginal one. The 
%original value of $\Rm$ multiplied with the artificial quenching factor evaluated with the total energy
%of the marginal solution then provides $\Rm_\tcrit$.

We used a spectral code based on the magnetic field's free decay modes, specified for vacuum boundary
conditions.
Each of these modes is characterized by a spherical harmonic $Y_l^m(\vartheta,\varphi)$ providing the angular dependence and a spherical Bessel function
$j_l(\mu_{ln}r)$ providing the radial one ($r,\vartheta,\varphi$ - spherical co--ordinates). Here,
$\mu_{ln}^2$ is 
the corresponding decay rate with $n$ giving roughly the number of zeros between center and surface of the
PNS.
  
Table \ref{dynprop} summarizes the critical magnetic Reynolds numbers $\Rm_\tcrit$
for the different velocities
and field types and some characteristic properties of the marginal magnetic eigensolutions,
 which are all non--oscillatory, that is,
$\Im\{\lambda\}=0$.
%Moreover, the
%maximum order of the spherical harmonics and Bessel functions contained in them is given together
%with the angular drift velocities of the fields in units of $1/t_{\text{Ohm}}$.

\begin{table}
\caption{\label{dynprop} Critical magnetic Reynolds numbers and characteristics of the marginal 
eigensolutions for the velocities considered.
$E_\tpol$ and $E_\ttor$ are the energies of the poloidal and toroidal parts of
the field, respectively, where the poloidal part comprises the externally observable multipole
constituents. The toroidal field is zero outside the PNS and its field lines lie completely
in spherical surfaces.
%$\omega_\tdrift$, given in units of $\eta/R^2$ is the angular drift
%velocity of the field with respect to the reference frame introduced in the caption of \figref{janv}
%to be understood in the sense of a rigid rotation of the field pattern.
$(l,n)_\tmax$ are the maximum spherical harmonic order and number of radial zeros, 
respectively, included in the numerical model; $(l,n)_\tdom$ correspond to the scales
dominating in the energy spectra of the solutions. }
\begin{center}
\begin{tabular}{l@{\hspace{6mm}}cc@{\hspace{9mm}}cc}
		      \hline					    
		      \hline					    \\[-.2cm]
                      &  \multicolumn{2}{c}{ 30 ms}     &      \multicolumn{2}{c}{0.9 s}          \\
		      &    A1    &   S1   &    A1    &   S1         \\
		      \hline					    \\
$\Rm_\tcrit$          &   4077	 &  4021  &  8720    &  6007        \\
$E_\tpol/E_\ttor$     &   0.24	 & 0.27   &  0.03    &  0.11        \\
$l_\tmax$    	      &    121 	 & 121    &  120     &  121         \\
$n_\tmax$    	      &     70 	 &  70    &   66     &   70         \\
$l_\tdom$             &     13   &  12    &    3     &   1          \\
$n_\tdom$             &      2   &   2    &    9     &   3         

\end{tabular}
\end{center}

\vspace{-.3cm}
\end{table}

\begin{figure*}
\hspace*{1.4cm}\begin{tabular}{@{\hspace{0.0\textwidth}}c@{\hspace{.1\textwidth}}c}
\epsfig{file=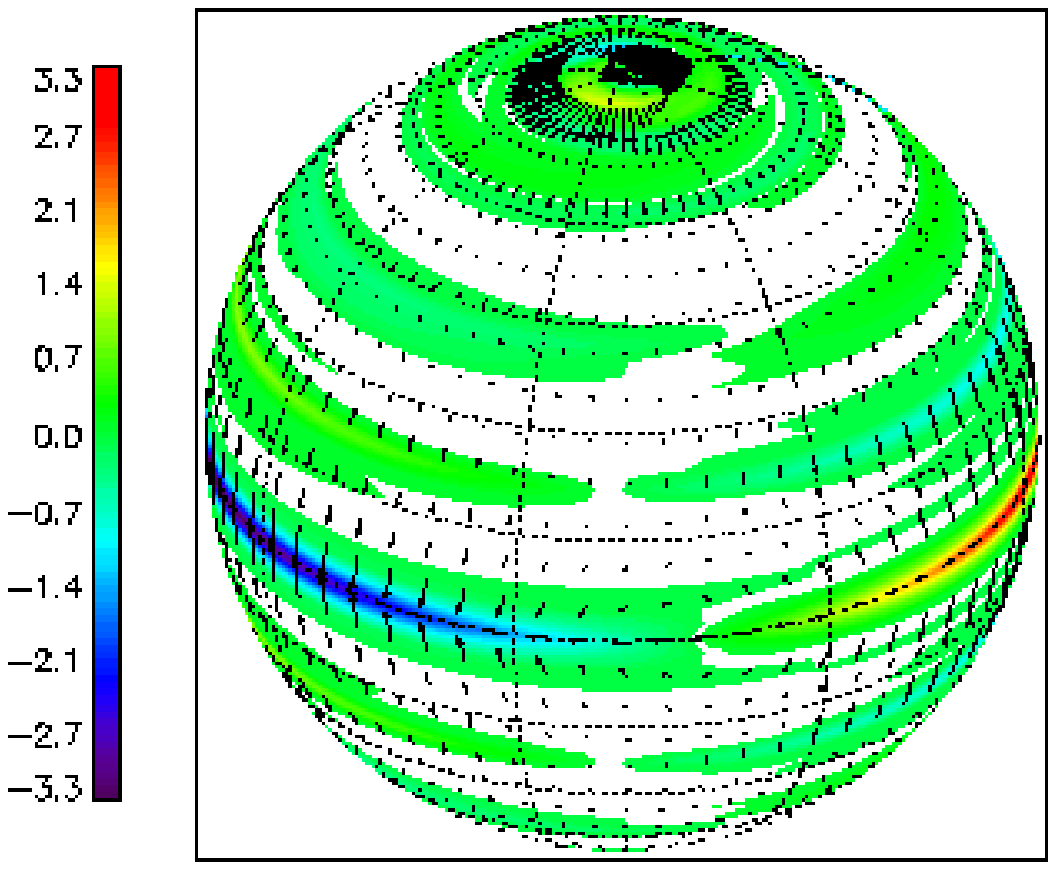,width = 0.369\textwidth}% Blickwinkel 30 60 
&
\epsfig{file=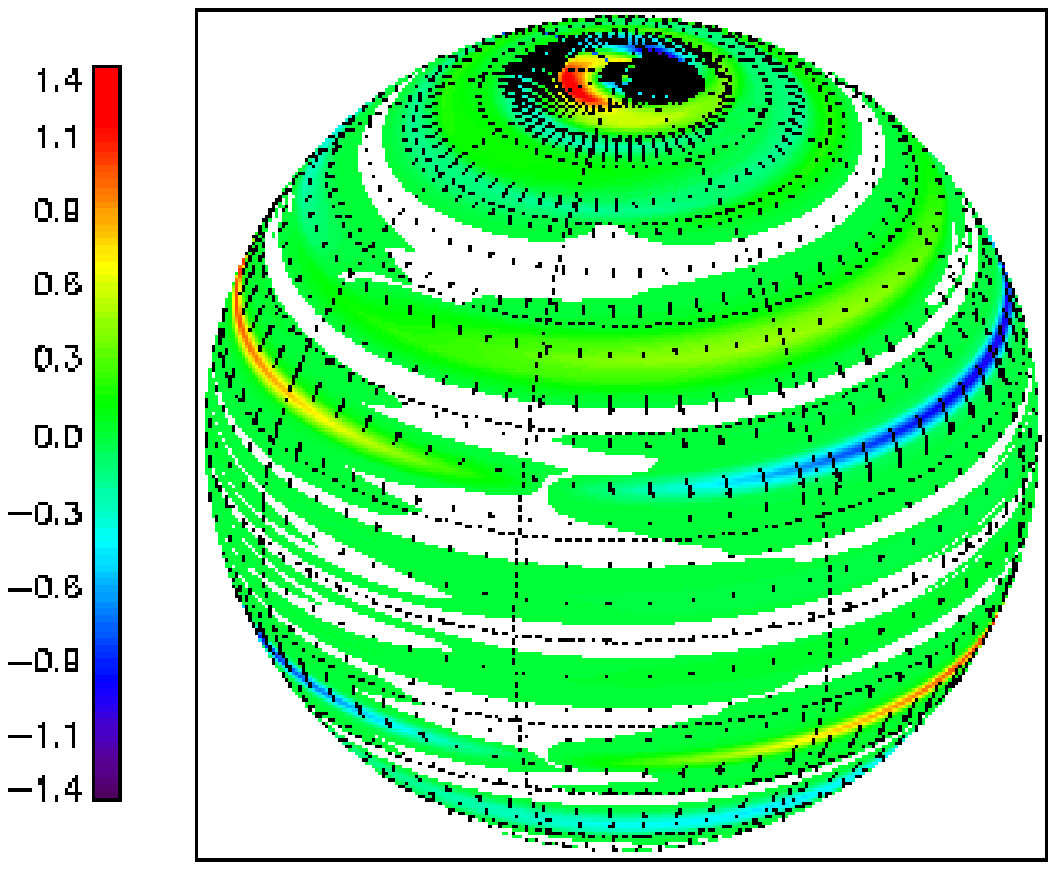,width = 0.369\textwidth}% Blickwinkel 30 -30
 \\*[2mm]
\epsfig{file=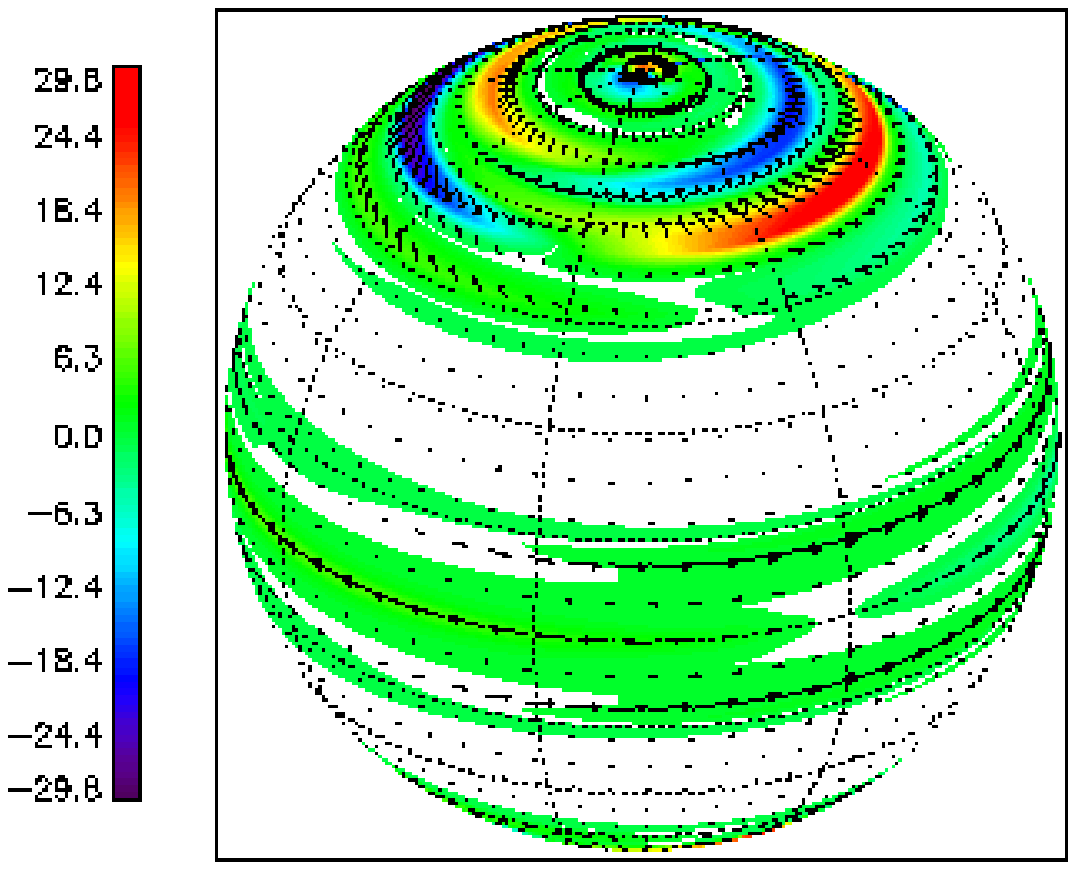,width = 0.369\textwidth}% Blickwinkel 30 60
&
\epsfig{file=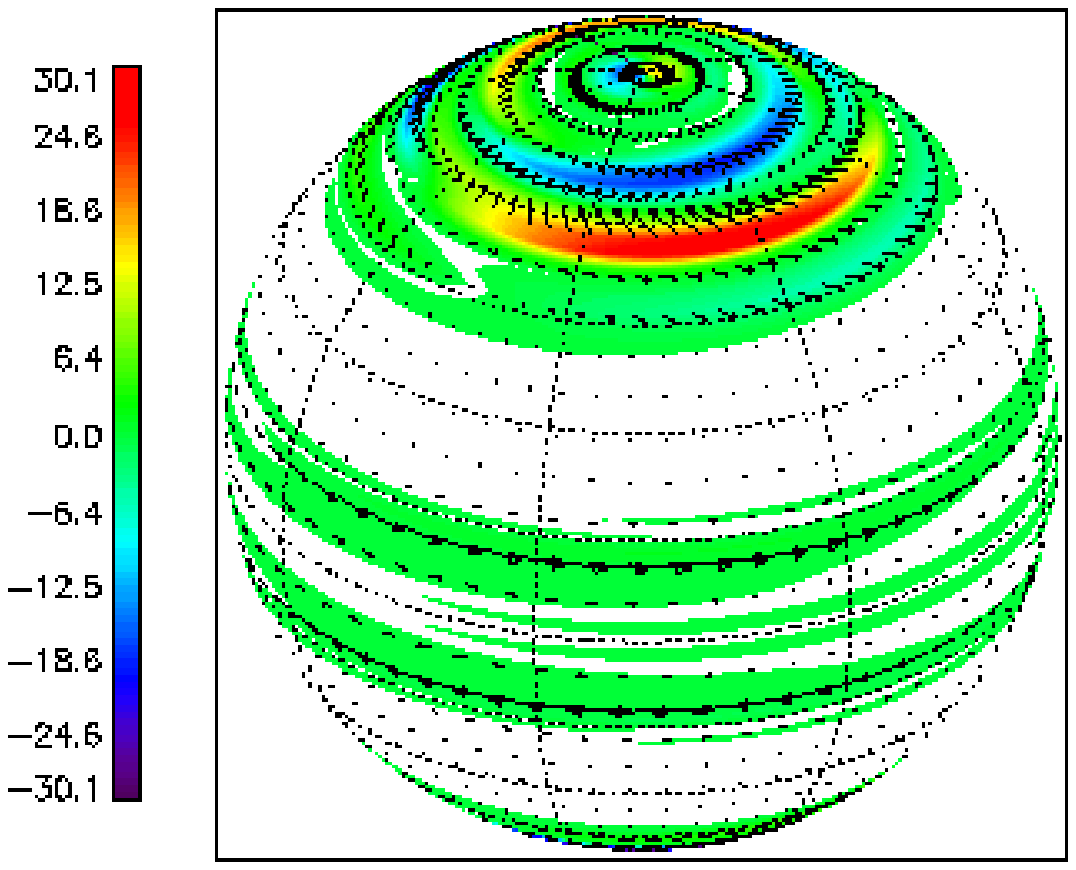,width = 0.369\textwidth}% Blickwinkel 30 -30
\end{tabular}
\caption{\label{jankadyn30ms} Marginal dynamo fields for the velocity from the
model of \cite{K97} at 30 ms after bounce. 
Left: S1, right A1 solution. Upper row: $r=R$, lower row $r=0.5 R$.
Arrows: tangential components, color encoding: normal component.
The field strengths are arbitrary as Eq. \eqref{indeqlin} is linear and homogeneous.
}
\end{figure*}

In \figsref{jankadyn30ms}{jankadyn1s} we present these magnetic fields on the surface of the PNS and on a
spherical surface inside the convective shell where field generation is supposed to be
most vigorous. Note that the A1 solutions contain quadrupoles, but those without  
any axis of symmetry ($Y_2^1(\vartheta,\varphi)$).
We emphasize that the magnetic fields shown belong to velocity fields extracted from 
the evolution of the PNS convection and then considered fixed, i.e., the fields corresponding to
the later moment has not evolved from the one corresponding to the earlier.

In their spherical harmonic spectra (referring to latitudinal scales), the `late' magnetic fields show somewhat larger scales than the
`early' ones,
but smaller scales in their spherical Bessel spectra (referring to radial scales).
Both S1 fields are at the surface predominant around the equator
where the early field has its maxima while the maxima of the late one occur
at $\pm 30^o$. 
%At least the appearance of the `early' fields can be understood from the corresponding velocity
%characterized mainly by two large cells which drag magnetic flux polewards in an outer shell of the PNS.
%Near the axis a second strong, but small cell inhibits the flux transport towards the equator and hence
%the closure of an otherwise possible cyclic transport.  
%Thus, the field is bound to be concentrated close to the axis.

%The `late' velocity and magnetic fields show features which let us speculate, 
%that in contrast to the early moment a mean--field scenario takes place [is realised].
%        
\enlargethispage{2\baselineskip}
\begin{figure*}
%\begin{center}
\hspace*{1.4cm}\begin{tabular}{@{\hspace{-0.0\textwidth}}c@{\hspace{.1\textwidth}}c}
\epsfig{file=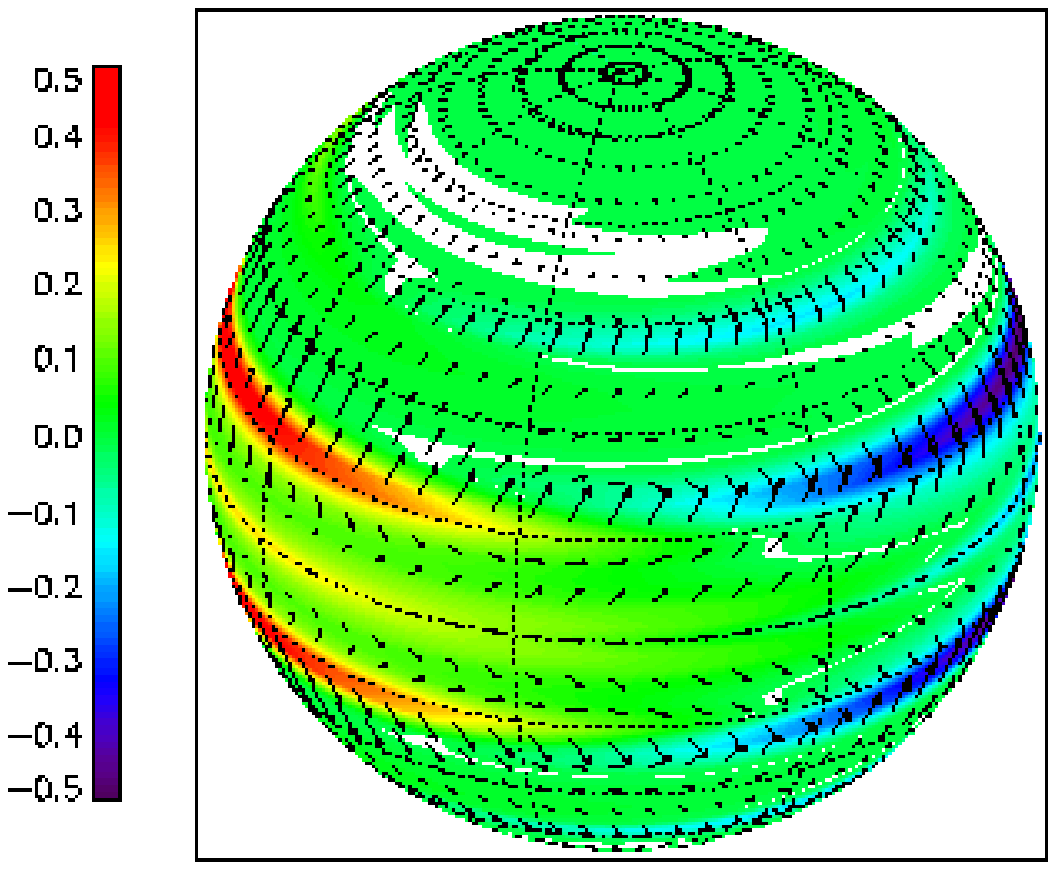,width = 0.369\textwidth}% Blickwinkel 30 -30
&
\epsfig{file=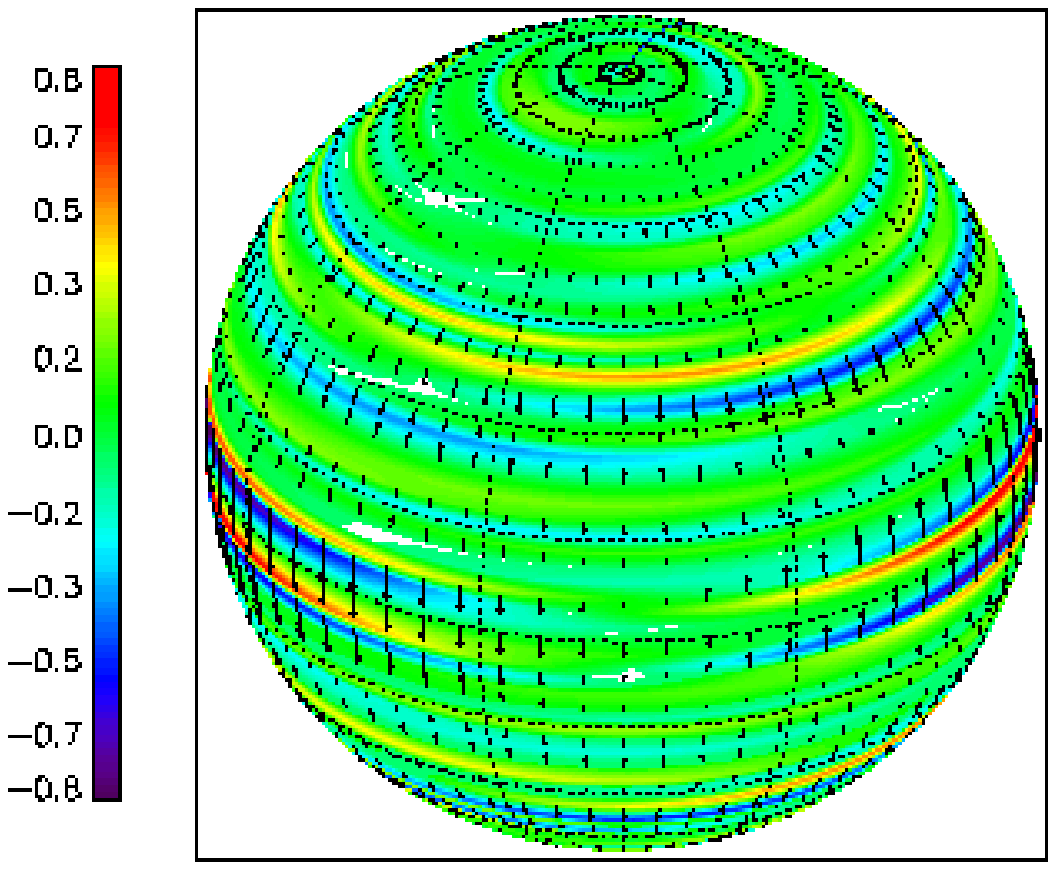,width = 0.369\textwidth}% Blickwinkel 30  20
\\*[2mm]
\epsfig{file=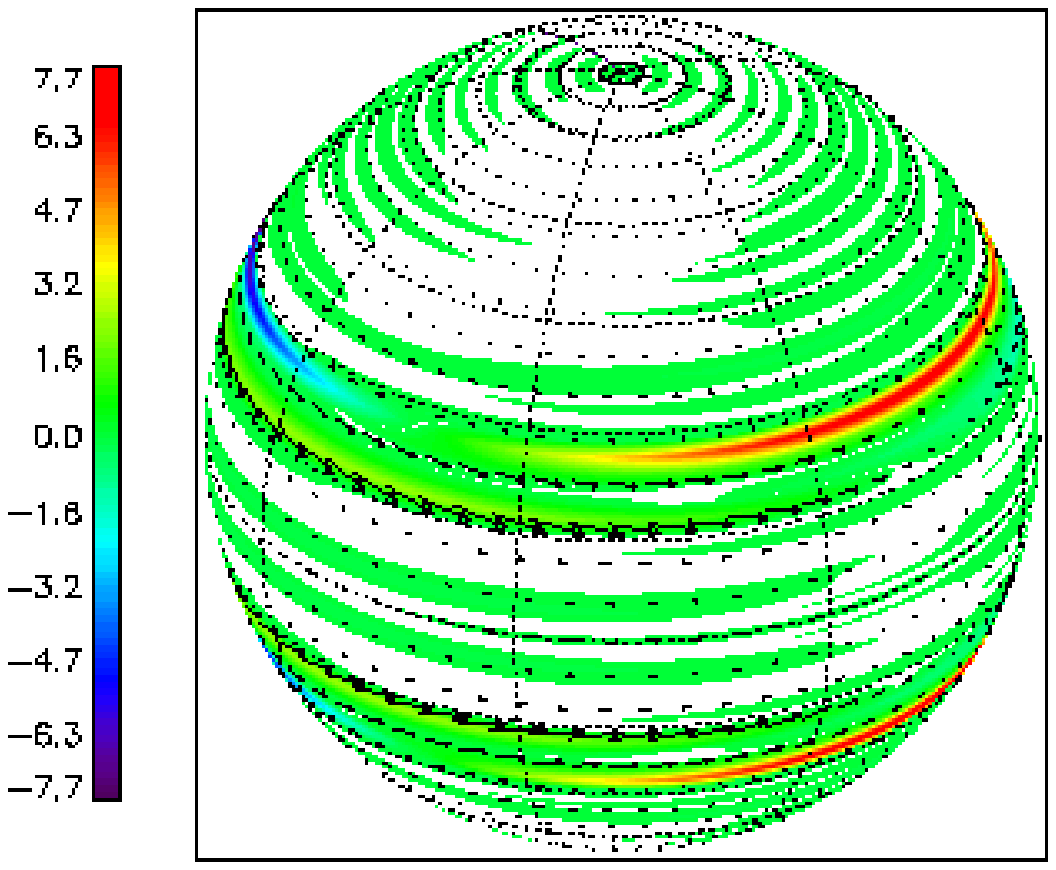,width = 0.369\textwidth}% Blickwinkel 30 -30
&
\hspace*{-3mm}\epsfig{file=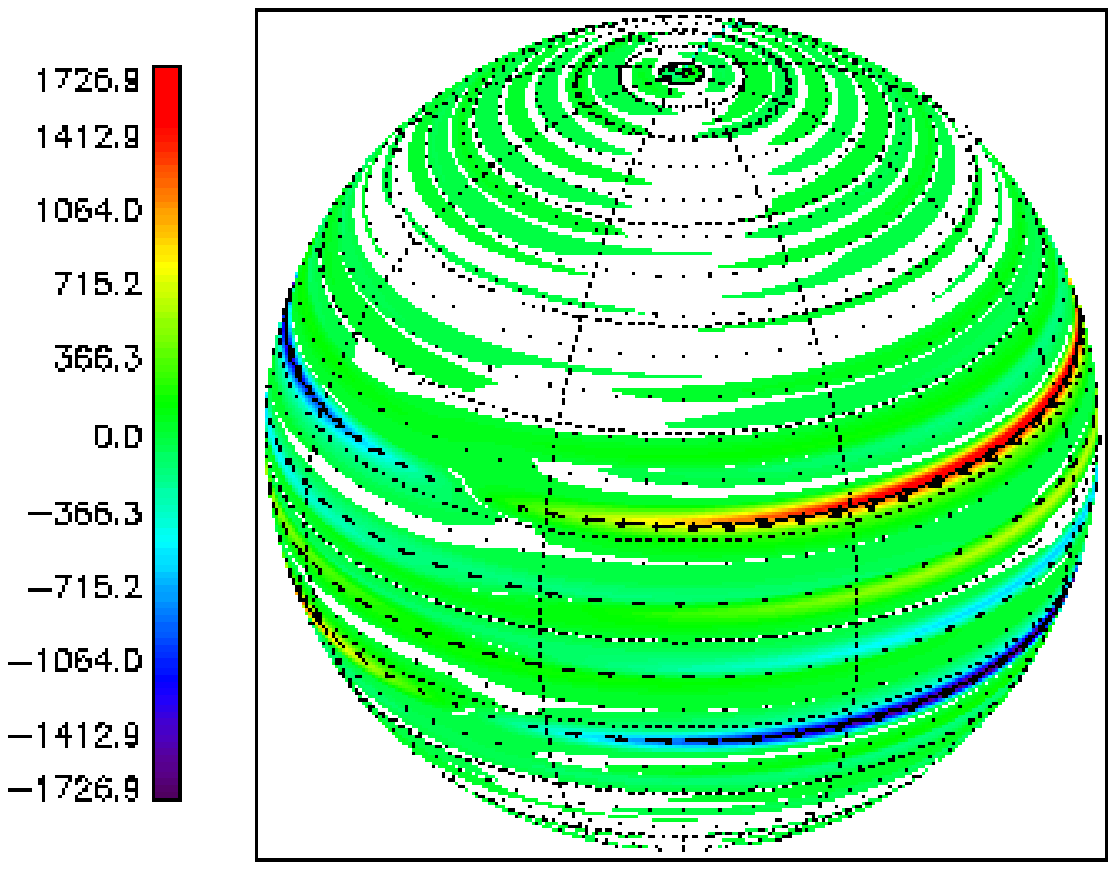,width = 0.39\textwidth}% Blickwinkel 30  20
\end{tabular}
%\end{center}
\caption{\label{jankadyn1s} The same as \figref{jankadyn30ms}, but for the
velocity at 0.9 s after bounce. Here, the radius of the inner surface (lower row) is  $r=0.6 R$.}
\end{figure*}
\section{Conclusions}
The main conclusion to be drawn is that the {\em geometrical structures} of the velocity fields employed are well suited to
amplify a magnetic seed field. However,
given the huge electric conductivity of the PNS matter ($\approx 10^{24}
\mbox{s}^{-1}$, see, e.g., \cite{TD93}, Table 1) and convective velocities of $\approx
10^8$cm s$^{-1}$, the magnetic Reynolds numbers turn out to be by about 16
orders of magnitude overcritical. 
As in general there is no monotonous relation between the growth rate of a 
dynamo and $\Rm$
nothing can thus be inferred with certainty about the dynamo activity of the considered
convective flows at their given strengths. 
However, it is known that due to increasing flux expulsion the growth rate as a function of $\Rm$ 
will typically reach a maximum and then decrease until the dynamo--ability 
of the flow may get lost completely or at least the growth rate approaches zero (`slow dynamo'). [See, e.g., \cite{Fetal97} for the Gubbins flow, where
the ratio of the upper to the lower critical $\Rm$ for the $m=1$ modes is only
5, or \cite{RRAF02} 
where it is shown that the $\alpha$ coefficient of the mean--field model of the Karlsruhe dynamo experiment
decreases with growing fluid velocity.]

There is no way to 
calculate magnetic field growth rates with the velocity amplitudes given above, 
as all experience shows that the field scales get smaller and smaller with 
growing $\Rm$. In our case, we are sure that with the available computing 
resources the necessary spatial resolution is by far unreachable.
Even if the problem would perhaps be less difficult in ideal MHD, one has to be aware that
then the Bondi--Gold theorem prohibits a kinematic dynamo solution unequal to
zero outside the PNS. 

On the other hand, PNS convection is a transient
phenomenon starting from small amplitudes and ending with the fluid at rest.
Let us discuss the possible scenarios assuming that the convection starts with 
a sufficiently small $\Rm$ and then passes, continuously growing, through a stage
which is critical with respect to the dynamo.

First we assume that
all stages of the convection beyond that point and up to 
a second critical stage only close to the very end of the convective 
phase show positive growth rates of the corresponding kinematic dynamos. 
Then, a magnetic field will be generated anyway (see \figref{schema}, case (i))
\footnote{We disregard here
the somewhat exotic possibility of a `suicidal' dynamo due to the existence
of two or more stable convective states which differ in their critical
magnetic Reynolds numbers. Then, in principle, the magnetic field could
switch an over-- into an undercritical state `killing' itself, see
\citet{FRR99}.}.

In a contrasting scenario, the increasing convection could leave its 
dynamo--active phase
very soon after having entered it (see \figref{schema}, case (ii)).
Then again two possibilities exist:
If the magnetic field grows so slowly that it cannot influence the 
flow significantly before it loses its dynamo--capability again, no strong
fields will be generated and the vigorous convection can even destroy
the (weak) field it had begun to build up before (see \figref{schema}, case
(iia)).
In the opposite case, the magnetic field grows so fast that its
Lorentz forces become able to hinder the convection from reaching amplitudes
too high for dynamo action (see \figref{schema}, case (iib)).
%In the first and the last cases discussed the influence of the generated 
%fields on the flow will obviously be essential, in the first case due to 
%the big growth rates
That is, the generated field could fundamentally 
change the flow in a very early stage.

In the first case of the second scenario, a further option is the recovery of 
the dynamo ability of the convection during its dying-out. Then, the period during
which the growth rate of the field is positive will be very short, limiting 
the accessible magnetic field strengths already on the kinematic level
(see \figref{schema}, case (iib)).

When accepting that the second scenario --- the convection is a `slow dynamo'
--- is the more probable one as
`fast dynamos' (growth rate positive and not approaching zero for
$\Rm\rightarrow\infty$) are rarely found, we must either
question the existence of an 
efficient PNS dynamo or declare that the influence of the generated 
magnetic field must not be neglected. Consequently, the possible 
essential role of PNS convection in enabling the supernova explosion must then 
be revised given its significant modification by the magnetic field at a very early
stage.
The only way to decide between the expounded alternatives 
is to perform MHD--calculations starting from the very beginning of the 
convection, i.e., with extremely small velocities.

We concede, that the two major scenarios (cases (i) and (ii) in \figref{schema})
represent only the extrema of an infinitude of possibilities which
could be constructed when assuming rapid changes in the dynamo--related properties
of the convection during its lifetime: Already for a fixed flow geometry
the dynamo growth rate $p$ as a function of $\Rm$  may change
its sign several times in the relevant range of values. But as the flow geometry is {\em not}
fixed during the lifetime of the convection, even its characterization as a slow or
fast dynamo could be different at different moments. 
Nevertheless, for any time interval with a positive $p$ terminated by a sign change
of $p$ the discussion summarized as cases (iia) and (iib) can be applied likewise leading again
to the above crucial alternative. Of course, possible consequences for the role of
PNS convection in supporting SN explosions can only be expected during (roughly) the first second
of the PNS's life.  

\begin{figure*}[t]
\input{baum}

\vspace{2mm}
\caption{\label{schema} Schematic representation of possible PNS dynamo scenarios.
$\Rm$ -- magnetic Reynolds number (solid and dotted [lower right panel]/green);
$p=\Re\{\lambda\}$ -- dynamo growth rate (dashed/red);
$\Emag$ -- energy of the entire PNS magnetic field (dash-dot/blue).
Colors refer to the online version.
Note that the curves for $\Rm$, $p$ and $\Emag$ are sketched as if no back--reaction
of the magnetic field on the convection existed (exception: dotted curve in lower right panel).}
\end{figure*}

\begin{acknowledgements} 
Thanks to T. Janka who provided us with the results of his and W. Keil's 
PNS simulations and to T. Janka, E. M\"uller, C. Fryer, and K.--H. R\"adler who discussed this paper
thoroughly with us. 

\end{acknowledgements} 
\bibliographystyle{aa}

\end{document}